\documentclass[12pt]{article}
\input epsf.sty
\usepackage{hyperref}

\usepackage{hyperref}
\usepackage{amssymb}
\usepackage{amsmath}
\usepackage[nosort]{cite}
\usepackage[normalem]{ulem}

\newcommand{\be}{\begin{equation}}
\newcommand{\ee}{\end{equation}}
\newcommand{\ben}{\begin{eqnarray}\displaystyle}
\newcommand{\een}{\end{eqnarray}}

\usepackage{amsmath}
\usepackage{amsfonts}
\usepackage{amssymb}
\usepackage{cite}
\usepackage{graphicx}

\def\sqr#1#2{{\vcenter{\vbox{\hrule height.#2pt
         \hbox{\vrule width.#2pt height#1pt \kern#1pt
            \vrule width.#2pt}
         \hrule height.#2pt}}}}

\usepackage{hyperref}
\hypersetup {
	colorlinks=true,
	linkcolor=blue,
	linktocpage=true,
	citecolor=blue,
	urlcolor=blue
}
\usepackage{geometry}
\usepackage{amsmath}
\usepackage{amsfonts}
\usepackage{amssymb}
\usepackage{cite}
\usepackage{graphicx}

\usepackage{setspace}
\begin{document}
\begin{center}
{
\Large{\bf  STRINGY INSTABILITY INSIDE \\\vspace{4mm} THE BLACK HOLE }}

\vspace{10mm}


\textit{ Nissan Itzhaki }
\break

Physics Department, Tel-Aviv University, \\
Ramat-Aviv, 69978, Israel \\
{\it nitzhaki@post.tau.ac.il}


\end{center}

\vspace{10mm}


\begin{abstract}
We show that negative $(\nabla \Phi)^2$, where $\Phi$ is the dilaton,  leads to a rapid creation
of folded strings.  Consequently it appears that the  interior of the $SL(2,\mathbb{R})_k/U(1)$ black hole is not empty, but is filled   with folded strings.
\end{abstract}


Motivated by \cite{Itzhaki:1996jt} we speculated sometime ago \cite{Itzhaki:2004dv} that a concrete way to express the challenge in having a non trivial structure at the horizon of a large Black Hole (BH) is to be  able to write down an effective action that renders the horizon special. We claimed that such an effective action must involve a "horizon order parameter"; an operator whose expectation value indicates if we are inside  or outside the BH. The horizon order parameter meant to be a trigger that modifies the physics inside the BH considerably compared to  the standard physics outside the BH.

Recently \cite{Itzhaki:2018rld} it was argued that in the case of the 2D $SL(2,\mathbb{R})_k/U(1)$ BH \cite{Elitzur:1991cb,Mandal:1991tz,Witten:1991yr}
 the horizon order parameter might take a particularly simple form
\be
{\cal O}=(\nabla \Phi)^2,
\ee
where $\Phi$ is the dilaton. Outside the $SL(2,\mathbb{R})_k/U(1)$ BH the operator ${\cal O}$ is positive while inside the BH it is negative. 

Some indirect evidence from the exact reflection coefficient of \cite{Teschner:1999ug} was provided   that the $SL(2,\mathbb{R})_k/U(1)$ BH interior is not empty in {\em classical} string theory \cite{interior,Itzhaki:2018rld}. These papers, however, did not explain how come the $SL(2,\mathbb{R})_k/U(1)$ BH is not empty or what it is filled with. Moreover no  relation with ${\cal O}$ was established. In particular it was not clear how the fact that ${\cal O}$  flips sign when crossing the horizon could possibly trigger non trivial effects inside the BH. 

In this short note we attempt  to fill up this gap. We argue that when ${\cal O}$ is negative  folded strings are created rapidly. As a result the $SL(2,\mathbb{R})_k/U(1)$ BH interior is not empty but is filled with folded strings. The idea that the BH is made out of fundamental strings is not new \cite{Susskind:1993ws,Horowitz:1996nw}. However, in the past these claims where made about small, stringy size, BHs while here we discuss  large BHs.

The classical background associated with the $SL(2,\mathbb{R})_k/U(1)$ BH is  (we work with $\alpha^{'}=1$)
\be\label{SL}
ds^2=-f(r)dt^2+\frac{dr^2}{f(r)},~~~~~\Phi(r)=\phi_0-Qr
\ee
where $f(r)=1-\mu e^{-2Qr}$. We focus on the supersymmetric case that describes the near horizon limit of $k$ NS5-branes in type II strings \cite{Maldacena:1997cg}. We consider the limit where, $Q=1/\sqrt{k}$, is small. In this limit the curvature is small away from the singularity. We also take $\phi_0\to -\infty$ so that stringy loops  can be neglected. The effect we describe happens already in classical string theory.

Away from the BH, when $r\to\infty$, we have  a 2D linear dilaton background 
\be\label{bacc}
ds^2=-dt^2+dr^2,~~~\Phi=-Qr,
\ee
with $Q>0$.
As far as the small fluctuations goes the length scale associated with the linear dilaton is $1/ Q$. This is evident from the background (\ref{bacc}) and also from the fact that the conformal dimension of an operator with momentum $p$ is shifted from $p^2/4$ to $p(p+Q)/4$. 

There are also classical long string solutions in this background \cite{Maldacena:2005hi}. An interesting aspect of these solutions, that will play a key role below, is that the relevant length scale associated with them is not $1 /Q$, but  $ Q$. 
Let us review their construction.  Consider a classical string that propagates non trivially in the background (\ref{bacc}). Working in the temporal gauge $t=\tau$ the equation of motions  are $\partial_{+}\partial_{-} r=0$ and the Virasoro constraints read
\be\label{c}
-1+(\partial_{+}r)^2+Q\partial^2_{+}r=0,~~~~~~~-1+(\partial_{-}r)^2+Q\partial^2_{-}r=0.
\ee
The solution to these equations is
\be\label{o}
r=r_0+Q \log\left( \frac12\left(\cosh\left(\frac{t-t_0}{Q}\right) +\cosh\left( \frac{\sigma}{Q} \right) \right) \right).
\ee
This solution is not periodic. It describes an infinity long folded string  that is stretched all the way to the weak coupling region. 

The turning point of the string, where it is at rest, is at $r=r_0$ and $t=t_0$. At $t$ much smaller or larger than $t_0$ the tip of the string is moving at a speed that is approaching the speed of light (see figure (1)). Note that unlike the yo-yo solution of \cite{Bardeen:1975gx,Bardeen:1976yt,Bars:1994qm} this solution is a smooth solution that does not involve discontinuities in $\partial_{\pm} r$.
\begin{figure}
\begin{center}
\includegraphics[scale=0.40]{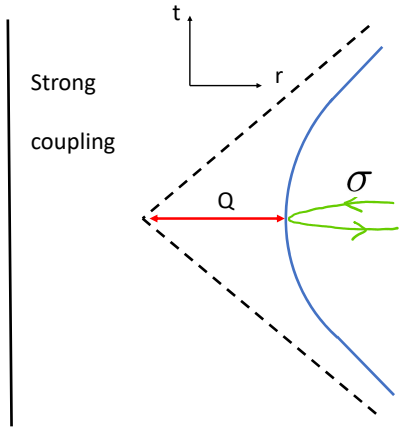}\vspace{-10mm}
\caption{The shape of the long folded string of \cite{Maldacena:2005hi}.  The green line indicates the way the string folds which is always  towards the weak coupling region. At early and late times the tip of the string is moving at a speed that approaches  $\pm 1$. }
\label{SBHpen}
\end{center}
\end{figure}

The total energy of such a string is infinite. The reason for the absent of finite energy classical string configurations  is that no matter how small $Q$ is the linear dilaton  prevents the string from folding towards the strong coupling.  
To see how this comes about it is instructive to consider (\ref{c}) at the turning point where the velocity vanishes and the Virasoro constraint are approximated by
\be\label{a}
\partial^2_{+}r=\frac{1}{Q},~~~~~~~\partial^2_{-}r=\frac{1}{Q}.
\ee
This implies that near the turning point there is a constant acceleration, that scales like $1/Q$. One way to think about this acceleration is that the slope of the dilaton induces a mass of the order of $Q$ at the tip of the string folds  \cite{Maldacena:2005hi}. Since we work in units in which the string tension is $1$ Newton's second law implies (\ref{a}) which leads to 
\be r=\frac{1}{2 Q}(\tau^2+\sigma^2).\ee
This shows that the string can fold only  in one direction, towards the weak coupling region.

The exact solution (\ref{o}) as well as the approximated solution (\ref{a})  imply that the length scale associated with the long string is $Q$ which for $Q\ll 1$ is much smaller than the perturbative scale, $1/Q$, associated with (\ref{bacc}). 
The fact that the non-perturbative long string introduces a new scale to the problem is closely related to the fact that the reflection coefficient in Liouville theory \cite{Zamolodchikov:1995aa} and in the $SL(2,\mathbb{R})_k$ model \cite{Teschner:1999ug} involve the non-perturbative momentum scale, $  1/Q$  ontop of the perturbative scale $Q$.

What happens when we approach the BH? Is it possible that outside the BH, where the curvature is of the order of $(\nabla\Phi)^2$ and (\ref{bacc}) is not a good approximation, there are finite energy classical configurations? There is a simple argument why, at least for small $Q$, the answer is no. To have a finite energy configuration the string should be able to turn both towards the BH and away from the BH. As discussed  above the length scale associated with the turning of the string  is $Q $. Curvature effects, that are absent in the discussion above, are important at much larger scales that are of the order of $1/Q$. Therefore, at least for small $Q$, the curvature associated with the solution (\ref{SL}) cannot change the conclusion that outside the BH the string can turn only towards the weak coupling region and hence there are only infinitely long string configurations outside the BH.

Next we wish to explore what happens inside the $SL(2,\mathbb{R})_k$ BH where  
$(\nabla\Phi)^2<0$. First we consider a useful toy model, that also have $(\nabla\Phi)^2<0$ - a time-like linear dilaton.
The background takes the form
\be\label{bac}
ds^2=-(dX^0)^2+(dX^1)^2,~~~\Phi=Q X^0.
\ee
Together with a matter CFT with a central charge such that the total central charge is 15 this is an exact background in classical superstring theory. The matter theory plays no role in  our discussion.  We take $Q>0$ so that, just like in the dynamically formed BH, the strong coupling is in the future.\footnote{When considering time-like linear dilaton as a toy model for cosmology, e.g. \cite{Hellerman:2006nx,Aharony:2006ra}, it is natural to have the strong coupling in the past.}

The gauge $X^0=\tau$ dismisses the effect of the linear dilaton. In this gauge the only solution, that does not involve discontinuous $\partial_{\pm} X^1$, is the trivial one $X^1=\sigma$. There are, however, non-trivial smooth solutions that are sensitive to the time-like linear dilaton. To reveal  them we work in the unusual gauge $X^1=\sigma$. 
Then the Virasoro constraints are
\be\label{cr}
1-(\partial_{+}X^0)^2-Q\partial^2_{+}X^0=0,~~~~~~~1-(\partial_{-}X^0)^2-Q\partial^2_{-}X^0=0,
\ee
and the solution is 
\be\label{s}
X^0=x^0+Q \log\left( \frac12\left( \cosh\left(\frac{X^1-x^1}{Q}\right)+\cosh\left(\frac{\tau}{Q}\right)\right)\right).
\ee
Despite the technical similarity with (\ref{o}) the physics associated with this solution is quite different than that of (\ref{o}). The string is created from the vacuum at a certain time $x^0$ and place $x^1$ and is folded towards the strong coupling (see figure (2)). At the tip the string is moving at an infinite speed that, for small $Q$, is quickly reduced towards the speed of light.  

\begin{figure}
\begin{center}
\includegraphics[scale=0.40]{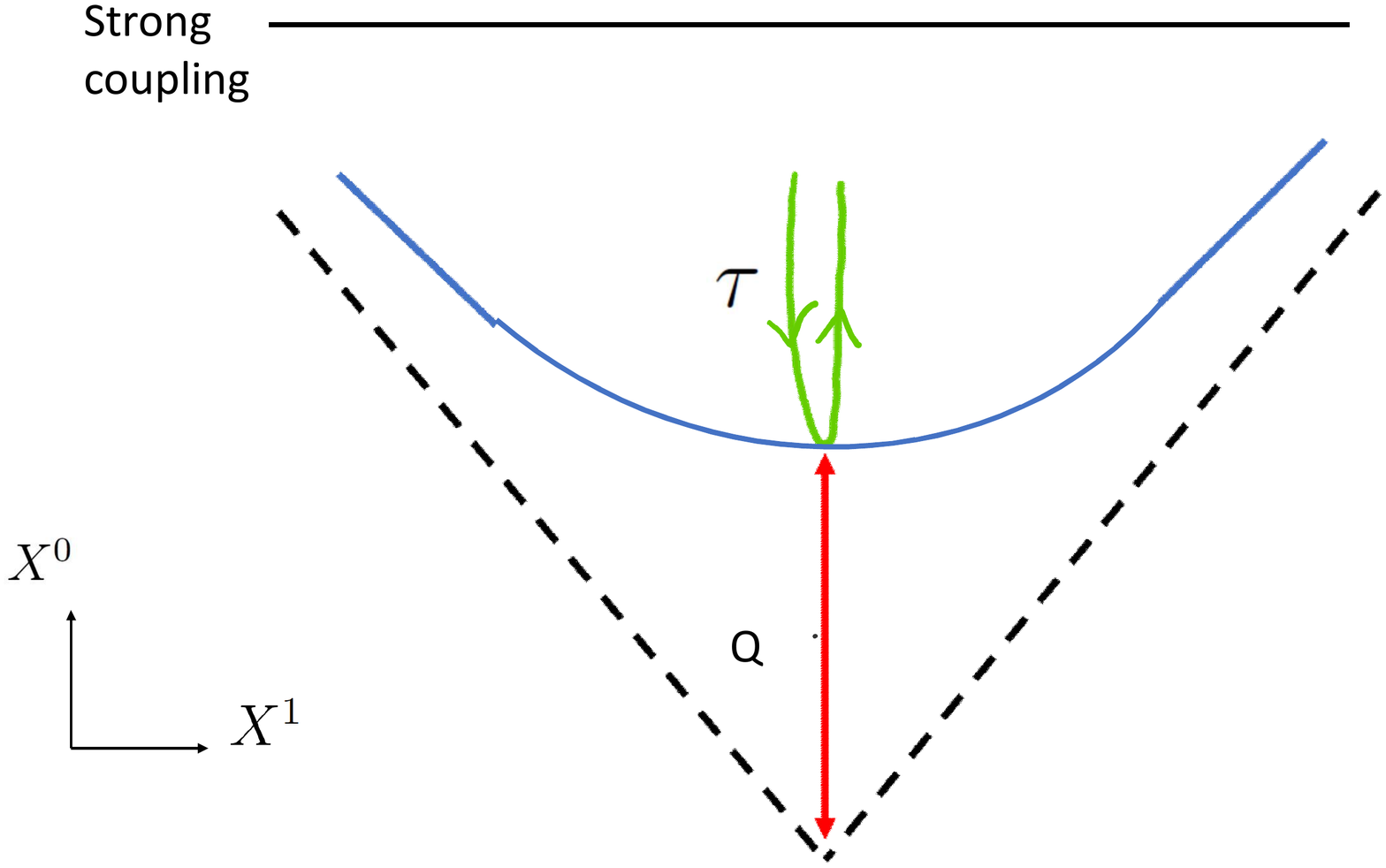}\vspace{-10mm}
\caption{The green line indicates the way the string folds which is always towards the strong coupling region.  The tip of the string is always moving faster than light. It approaches the speed of light, the dashed lines, when  $|X_1|$ is large.}
\label{SBHpen}
\end{center}
\end{figure}

At first sight, figure (2), that describes (\ref{s}),  appears to resemble figure (3) that describes the Schwinger mechanism \cite{Schwinger:1951nm} or more precisely its stringy generalization (see e.g. \cite{Dowker:1995sg}). There are, however, crucial differences. First, (\ref{s}) is a {\it classical} solution in Minkowski signature. A  semi-classical description of  the Schwinger mechanism that starts from the vacuum and ends with on-shell particles involves gluing a Minkowskian solution with a Euclidean solution (see figure 3). The Euclidean section implies that, unlike in (\ref{s}), the Schwinger mechanism is a quantum process that involves tunneling and is exponentially suppressed by $S_{E}/\hbar.$ The lack of a Euclidean section in (\ref{s}) implies that  the creation rate associated with it is not expected to be suppressed exponentially.

Second, in the Schwinger mechanism the particles (or strings) are moving slower than light. This is not the case in (\ref{s}). The points where the string folds are moving faster than light. This suggests the following as the mechanism behind  this solution.
 In analogy with \cite{Maldacena:2005hi} the  linear dilaton  induces a  mass at the point where the string folds. Only that now since we have a time-like linear dilaton the mass is tachyonic, $m_{tip}^2\sim-Q^2$. Condensation of this tachyon  attempts to generate a runaway behaviour. It is the tension of the string that holds the configuration together. The balance between the two  is what drives  (\ref{s}) and induces the large momentum scale, $1/Q$, associated with it.

We do not know what is the end-point of the condensation of this classical configuration. It would be particularly interesting if it leads to a resolution of the future strong coupling singularity.

\begin{figure}
\begin{center}
\includegraphics[scale=0.45]{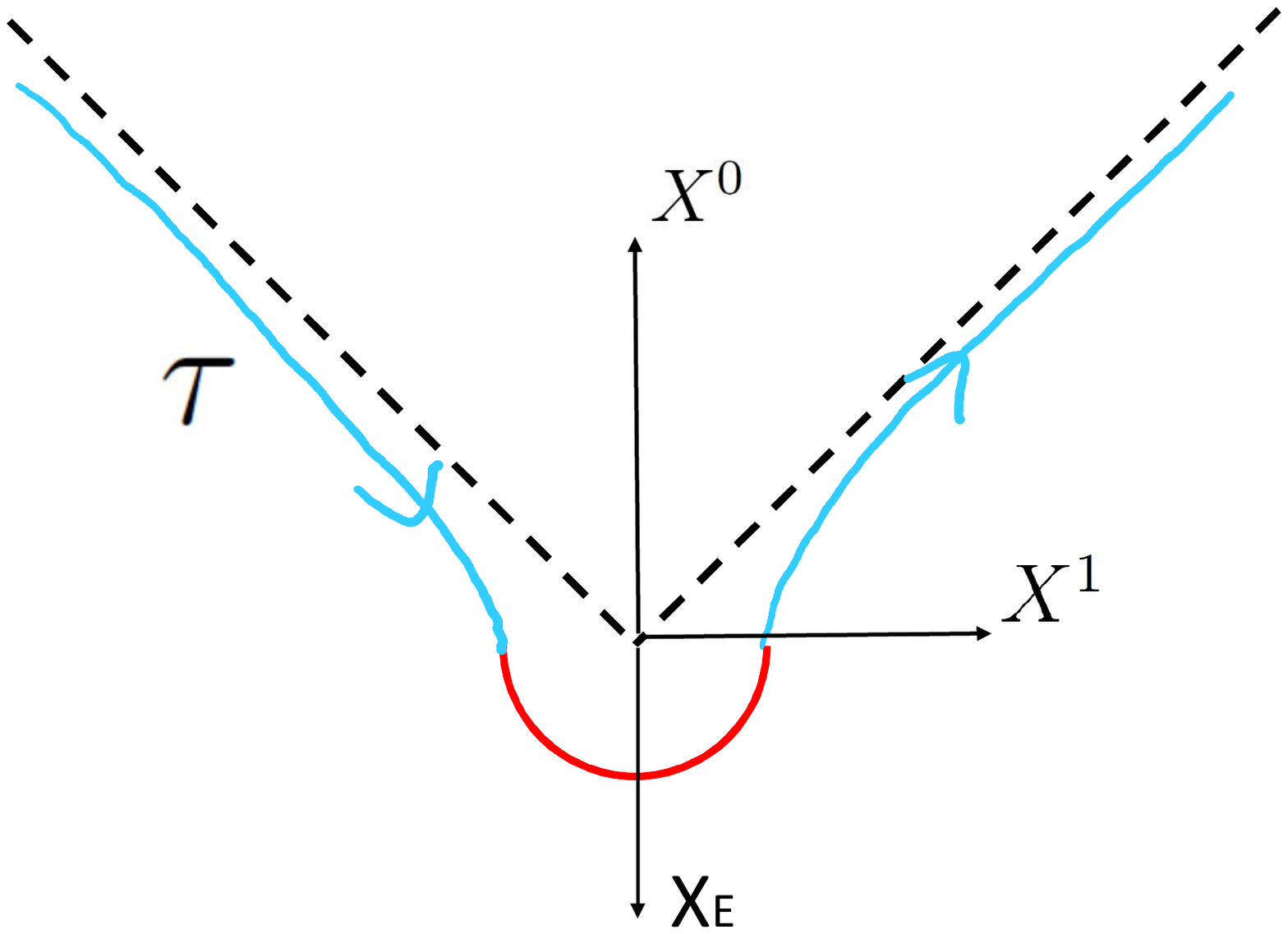}\vspace{-10mm}
\caption{The Schwinger mechanism. The red line represents the Euclidean solution. It is glued at $X_E=X^0=0$ to the Minkowskian solution that is represented by the blue lines. }
\label{SBHpen}
\end{center}
\end{figure}

We are now in a position to describe the folded string inside the $SL(2,\mathbb{R})_k/U(1)$ BH.
The slope of the time-like dilaton inside the $SL(2,\mathbb{R})_k/U(1)$ BH is not constant and  the metric is not flat. Hence (\ref{bac}) does not describe the region behind the horizon. However, $(\nabla\Phi)^2$ is negative there and  we saw that the length scale associated with the folded string creation is $Q$. This implies that inside the $SL(2,\mathbb{R})_k/U(1)$ BH the length scale associated with the creation of the folded string  is $\sqrt{-(\nabla\Phi)^2}$. This scale is typically of order $1/\sqrt{k}$ which is much smaller than the curvature scale $\sqrt{k}$. 
 Hence in the large $k$ limit the folded string creation is, for all practical purposes, a local process.  This means that the BH curvature  cannot modify much the creation of the folded string. It will surely modify  (\ref{s}) at distances of order $\sqrt{k}$,
but not the conclusion that such folded strings are created. We expect a typical folded string inside the $SL(2,\mathbb{R})_k/U(1)$ BH to take the shaped presented in figure (4).  

The folded string condensation will not stop until $(\nabla\Phi)^2$ is not negative everywhere inside the BH. This means that the entire BH interior should be filled with folded strings. This appears to be the case both for eternal and dynamically formed $SL(2,\mathbb{R})_k/U(1)$ BH.

\begin{figure}
\begin{center}
\includegraphics[scale=0.40]{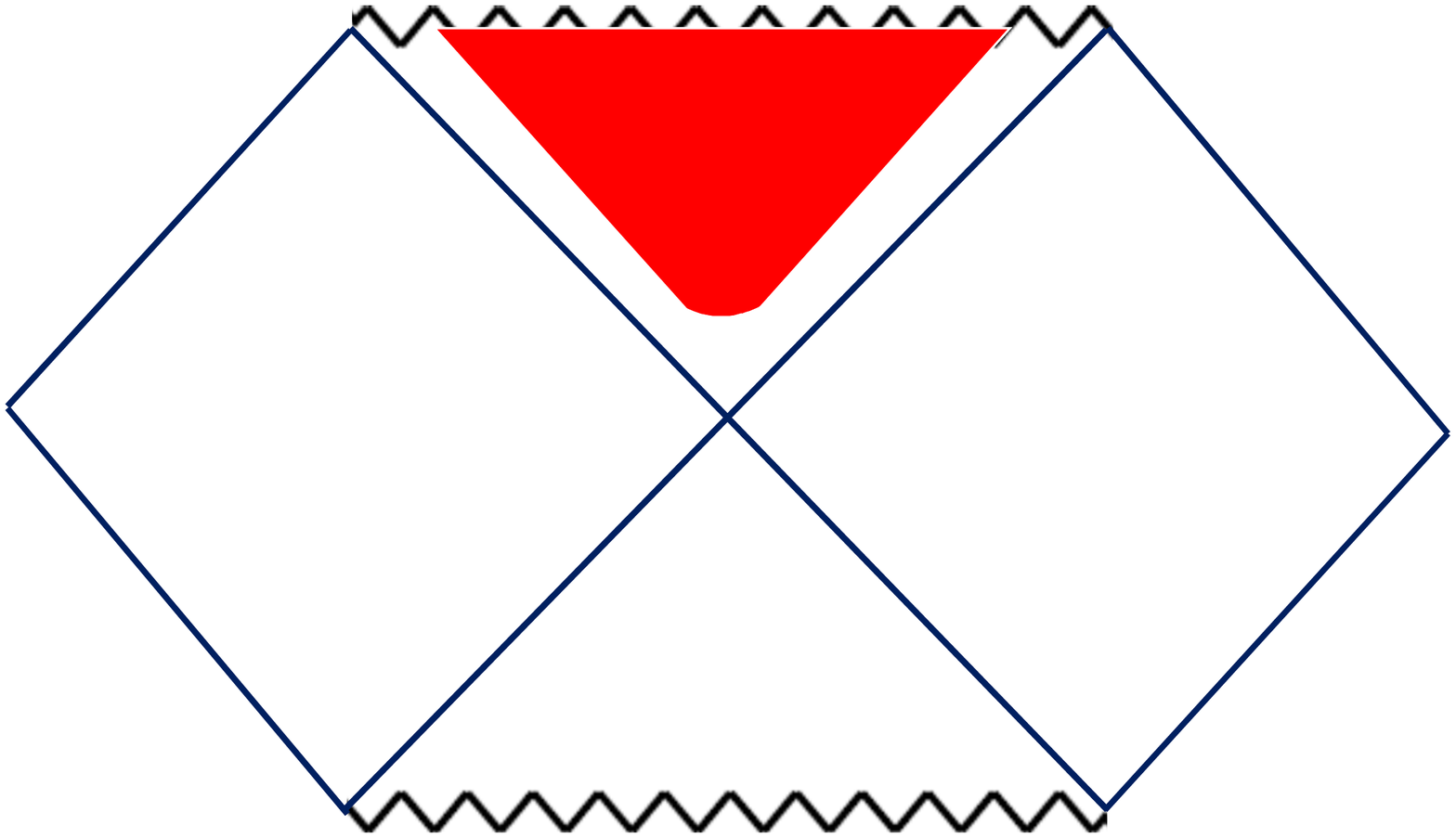}
\vspace{-10mm}
\caption{A typical  folded string inside the $SL(2,\mathbb{R})_k/U(1)$  BH. }
\label{SBHpen}
\end{center}
\end{figure}

This  conclusion seems to fit neatly with claims made in  \cite{interior,Itzhaki:2018rld} that the potential associated with the $SL(2,\mathbb{R})_k/U(1)$ BH blows up just behind the horizon. 
The unusual properties of (\ref{s}) suggests that its backreaction 
is likely to be non-standard and could drastically modify the BH interior. It remains to be seen if precise connection with the potential found in 
 \cite{interior,Itzhaki:2018rld} can be made. The fact that the potential behind the horizon found in \cite{interior,Itzhaki:2018rld} is determined by the same length scale, $1/\sqrt{k}$, that controls (\ref{s}) seems to support this.

The conclusion that the interior of the eternal $SL(2,\mathbb{R})_k/U(1)$ BH is filled with strings also seems to go well with some  Euclidean reasoning. The FZZ duality \cite{fzz,Kazakov:2000pm} implies that at the tip of  the cigar there is a condensation of a winding one tachyon mode \cite{Kutasov:2000jp,juan}. The Harlte Hawking wave function \cite{Hartle:1976tp} then seems to imply that the BH interior should be filled with strings  (see figure 5). It seems reasonable to suspect that the wave function of the winding one tachyon mode found in \cite{Giveon:2013ica} is related to the wave function of the folding point  of the string discussed here. One needs to go beyond the classical discussion presented here to study this.

\begin{figure}
\centerline{\includegraphics[scale=0.5]{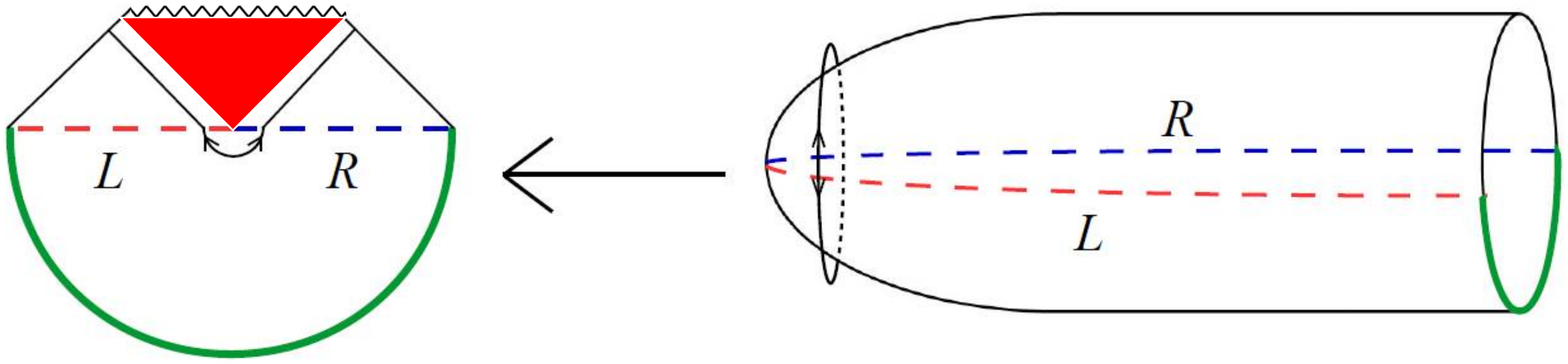}}
\vspace{-20mm}
\caption{The winding one tachyon mode that condenses at the tip of the cigar appears to be related to the folded string that fills up the entire $SL(2,\mathbb{R})_k/U(1)$ BH interior.}
\label{winding}
\end{figure}

The $SL(2,\mathbb{R})_k/U(1)$ BH is the near horizon of $k$ NS5-branes \cite{Maldacena:1997cg}. The fundamental string is the electric-magnetic dual of the NS5-brane \cite{Callan:1991at}. It is, therefore, natural to speculate that the region behind the horizon of near extremal Dp-branes is filled with their electric-magnetic dual, D(6-p)-branes \cite{Polchinski:1995mt}.

In this note we focused on BHs. It is likely, however, that the folded string creation could have interesting applications to cosmology as well.

\section*{Acknowledgments}

We thank K. Attali, A. Giveon and L. Liram for discussions.
This work  is supported in part by the I-CORE Program of the Planning and Budgeting Committee and the Israel Science Foundation (Center No. 1937/12), and by a center of excellence supported by the Israel Science Foundation (grant number 1989/14).


\begin{thebibliography}{1}

\bibitem{Itzhaki:1996jt} 
  N.~Itzhaki,
  ``Is the black hole complementarity principle really necessary?,''
  hep-th/9607028.

\bibitem{Itzhaki:2004dv} 
  N.~Itzhaki,
  ``The Horizon order parameter,''
  hep-th/0403153.

\bibitem{Itzhaki:2018rld} 
  N.~Itzhaki and L.~Liram,
  ``A stringy glimpse into the black hole horizon,''
  JHEP {\bf 1804}, 018 (2018)
  doi:10.1007/JHEP04(2018)018
  [arXiv:1801.04939 [hep-th]].

\bibitem{Elitzur:1991cb}
  S.~Elitzur, A.~Forge and E.~Rabinovici,
  ``Some global aspects of string compactifications,''
  Nucl.\ Phys.\ B {\bf 359} (1991) 581.



\bibitem{Mandal:1991tz}
  G.~Mandal, A.~M.~Sengupta and S.~R.~Wadia,
  ``Classical solutions of two-dimensional string theory,''
  Mod.\ Phys.\ Lett.\ A {\bf 6} (1991) 1685.

\bibitem{Witten:1991yr}
  E.~Witten,
  ``On string theory and black holes,''
  Phys.\ Rev.\ D {\bf 44}, 314 (1991).


\bibitem{Teschner:1999ug} 
  J.~Teschner,
  ``Operator product expansion and factorization in the H+(3) WZNW model,''
  Nucl.\ Phys.\ B {\bf 571}, 555 (2000)
  doi:10.1016/S0550-3213(99)00785-3
  [hep-th/9906215].




\bibitem{interior}
  R.~Ben-Israel, A.~Giveon, N.~Itzhaki and L.~Liram,
  ``On the black hole interior in string theory,''
  JHEP {\bf 1705}, 094 (2017)
  [arXiv:1702.03583 [hep-th]].
 
\bibitem{Susskind:1993ws} 
  L.~Susskind,
  ``Some speculations about black hole entropy in string theory,''
  In *Teitelboim, C. (ed.): The black hole* 118-131
  [hep-th/9309145].

\bibitem{Horowitz:1996nw} 
  G.~T.~Horowitz and J.~Polchinski,
  ``A Correspondence principle for black holes and strings,''
  Phys.\ Rev.\ D {\bf 55}, 6189 (1997)
  doi:10.1103/PhysRevD.55.6189
  [hep-th/9612146].

\bibitem{Maldacena:2005hi} 
  J.~M.~Maldacena,
  ``Long strings in two dimensional string theory and non-singlets in the matrix model,''
  JHEP {\bf 0509}, 078 (2005)
  [Int.\ J.\ Geom.\ Meth.\ Mod.\ Phys.\  {\bf 3}, 1 (2006)]
  doi:10.1088/1126-6708/2005/09/078, 10.1142/S0219887806001053
  [hep-th/0503112].



\bibitem{Bardeen:1975gx} 
  W.~A.~Bardeen, I.~Bars, A.~J.~Hanson and R.~D.~Peccei,
  ``A Study of the Longitudinal Kink Modes of the String,''
  Phys.\ Rev.\ D {\bf 13}, 2364 (1976).
  doi:10.1103/PhysRevD.13.2364
\bibitem{Bardeen:1976yt} 
  W.~A.~Bardeen, I.~Bars, A.~J.~Hanson and R.~D.~Peccei,
  ``Quantum Poincare Covariance of the D = 2 String,''
  Phys.\ Rev.\ D {\bf 14}, 2193 (1976).
  doi:10.1103/PhysRevD.14.2193
\bibitem{Bars:1994qm} 
  I.~Bars,
  ``Folded strings,''
  Lect.\ Notes Phys.\  {\bf 447}, 26 (1995)
  [hep-th/9412044].



\bibitem{Zamolodchikov:1995aa} 
  A.~B.~Zamolodchikov and A.~B.~Zamolodchikov,
  ``Structure constants and conformal bootstrap in Liouville field theory,''
  Nucl.\ Phys.\ B {\bf 477}, 577 (1996)
  doi:10.1016/0550-3213(96)00351-3
  [hep-th/9506136].

\bibitem{Hellerman:2006nx} 
  S.~Hellerman and I.~Swanson,
  ``Cosmological solutions of supercritical string theory,''
  Phys.\ Rev.\ D {\bf 77}, 126011 (2008)
  doi:10.1103/PhysRevD.77.126011
  [hep-th/0611317].

\bibitem{Aharony:2006ra} 
  O.~Aharony and E.~Silverstein,
  ``Supercritical stability, transitions and (pseudo)tachyons,''
  Phys.\ Rev.\ D {\bf 75}, 046003 (2007)
  doi:10.1103/PhysRevD.75.046003
  [hep-th/0612031].

\bibitem{Schwinger:1951nm} 
  J.~S.~Schwinger,
  Phys.\ Rev.\  {\bf 82}, 664 (1951).
  doi:10.1103/PhysRev.82.664

\bibitem{Dowker:1995sg} 
  F.~Dowker, J.~P.~Gauntlett, G.~W.~Gibbons and G.~T.~Horowitz,
  ``Nucleation of p-branes and fundamental strings,''
  Phys.\ Rev.\ D {\bf 53}, 7115 (1996)
  doi:10.1103/PhysRevD.53.7115
  [hep-th/9512154].

\bibitem{fzz} V. A. Fateev, A. B. Zamolodchikov and Al. B. Zamolodchikov, unpublished.

\bibitem{Kazakov:2000pm} 
  V.~Kazakov, I.~K.~Kostov and D.~Kutasov,
  ``A Matrix model for the two-dimensional black hole,''
  Nucl.\ Phys.\ B {\bf 622}, 141 (2002)
  doi:10.1016/S0550-3213(01)00606-X
  [hep-th/0101011].

\bibitem{Kutasov:2000jp} 
  D.~Kutasov and D.~A.~Sahakyan,
  ``Comments on the thermodynamics of little string theory,''
  JHEP {\bf 0102}, 021 (2001)
  doi:10.1088/1126-6708/2001/02/021
  [hep-th/0012258].

\bibitem{juan}
J.~Maldacena, talk at Strings 2004, Paris. 
  
\bibitem{Hartle:1976tp} 
  J.~B.~Hartle and S.~W.~Hawking,
  ``Path Integral Derivation of Black Hole Radiance,''
  Phys.\ Rev.\ D {\bf 13}, 2188 (1976).
  doi:10.1103/PhysRevD.13.2188

\bibitem{Giveon:2013ica} 
  A.~Giveon and N.~Itzhaki,
  ``String theory at the tip of the cigar,''
  JHEP {\bf 1309}, 079 (2013)
  doi:10.1007/JHEP09(2013)079
  [arXiv:1305.4799 [hep-th]].
  
\bibitem{Maldacena:1997cg} 
  J.~M.~Maldacena and A.~Strominger,
  ``Semiclassical decay of near extremal five-branes,''
  JHEP {\bf 9712}, 008 (1997)
  doi:10.1088/1126-6708/1997/12/008
  [hep-th/9710014].



\bibitem{Callan:1991at} 
  C.~G.~Callan, Jr., J.~A.~Harvey and A.~Strominger,
  ``Supersymmetric string solitons,''
  In *Trieste 1991, Proceedings, String theory and quantum gravity '91* 208-244 and Chicago Univ. - EFI 91-066 (91/11,rec.Feb.92) 42 p
  [hep-th/9112030].

\bibitem{Polchinski:1995mt} 
  J.~Polchinski,
  ``Dirichlet Branes and Ramond-Ramond charges,''
  Phys.\ Rev.\ Lett.\  {\bf 75}, 4724 (1995)
  doi:10.1103/PhysRevLett.75.4724
  [hep-th/9510017].


\end{thebibliography}
\end{document}